# *Flimma*: A federated and privacy-preserving tool for differential gene expression analysis

Olga Zolotareva[1*], Reza Nasirigerdeh[1*], Julian Matschinske[1], Reihaneh Torkzadehmahani[1], Tobias Frisch[2], Julian Späth[1], David B. Blumenthal[1], Amir Abbasinejad[1,4], Paolo Tieri[3,4], Nina K. Wenke[1], Markus List[1], Jan Baumbach[1,2]

1 Chair of Experimental Bioinformatics, TUM School of Life Sciences, Technical University of Munich, Munich, Germany
2 Department of Mathematics and Computer Science, University of Southern Denmark, Odense, Denmark
3 CNR National Research Council, IAC Institute for Applied Computing, Rome, Italy
4 SapienzaUniversity of Rome, Rome, Italy
* Joint first authors

## Abstract

Aggregating transcriptomics data across hospitals can increase sensitivity and robustness of differential expression analyses, yielding deeper clinical insights. As data exchange is often restricted by privacy legislation, meta-analyses are frequently employed to pool local results. However, if class labels are inhomogeneously distributed between cohorts, their accuracy may drop.
*Flimma* (https://exbio.wzw.tum.de/flimma/) addresses this issue by implementing the state-of-the-art workflow *limma voom* in a privacy-preserving manner, i.e. patient data never leaves its source site. *Flimma* results are identical to those generated by *limma voom* on combined datasets even in imbalanced scenarios where meta-analysis approaches fail.

## Background

The identification of differentially expressed genes or transcripts, e.g. in diseases or in response to treatment, is a standard but important task in molecular systems medicine.

Differential gene expression analysis compares the expression profiles of two or more groups of samples to reveal genes with significant differences between the groups. Technologies for high-throughput gene expression profiling include microarrays and RNA sequencing, the latter being more widely used in clinical research today. Both are intrinsically different, e.g. signal- vs. count-based measurement, and their results subject to platform-specific biases [1,2]. Many bioinformatics tools for identifying differentially expressed genes from such data have been developed [3–9]. These methods differ with respect to the assumptions about data distribution (e.g. normal vs. Poisson or negative binomial distribution), the data normalization strategies, and in the test statistic used to detect differentially expressed genes [10–12].

One major challenge of differential expression studies is the lack of robustness due to the high technical and biological variability of the data [13–15], which can be addressed using various strategies [16–18]. The simplest and most effective way would be to increase the sample size [16], which is non-trivial, as data collection is expensive and time-consuming, sample availability may be limited (e.g. metastatic cancer or healthy tissue samples are difficult to obtain), or because existing data can not be shared and pooled as they are subject to personal data protection laws. The latter is of particular concern for next-generation sequencing data, from which the sample donor can be identified under certain conditions [19–21]. Although several human-derived expression profiles are nowadays publicly available, their utility (in particular in clinical settings) is often still limited for inherent privacy issues. Recent works suggest that patient genotypes can be predicted from RNA-seq data, making patients identifiable through expression profiles or eQTL data obtained from open-access sources [22–24]. In addition, publicly available expression data often lacks relevant clinical metadata that is important for the statistical analysis and thus may require harmonization or even filtering prior to pooling.

To control the exchange of sensitive molecular profiling data from e.g. next-generation sequencing experiments, databases, such as dbGaP [25] or EGA [26], restrict access to authorized users affiliated with organizations willing to guarantee the legal and secure use of personal data. Nevertheless, the application procedure needs to be repeated per study and per database, making this a difficult and time-consuming process, which is also error-prone if *a priori* unknown confounder variables are not requested and can thus not be corrected for in the downstream analysis. Alternatively, when direct access to raw data is not possible, researchers can combine the results of several studies using meta-analysis techniques such as Fisher's method [27], Stouffer's method [28], RankProd [29], or the random effects model [30] (REM). Meta-analysis is widely adopted for aggregation of genome-wide association studies (GWAS) [31] and differential gene expression analysis results [32,33] (cf. "Meta-analysis approaches" in the "Methods" section for details). The main disadvantage of meta-analysis tools is that their underlying assumptions about the distribution of p-values or effect sizes may not be realistic. Furthermore, meta-analysis largely ignores possible differences between cohorts (e.g., class imbalance or heterogeneity of covariate distributions) [34] or data processing steps (e.g., normalization) [35], which may have a significant impact on the results [34].

Privacy-preserving techniques, such as federated learning, have recently moved into the focus of research for tasks involving privacy-sensitive patient data. Federated learning has become increasingly popular in bioinformatics for GWAS [36,37], survival analysis [38], and additional challenges in patient data processing [39,40]. Federated learning implies collaborative model training by multiple participants without disclosing private data to any other party [41]. Instead, each participant only shares intermediate model parameters while keeping the private data in the local environment (e.g. the legally safe harbors of the hospitals' IT system). Only model parameters, which are by design privacy-preserving, are communicated and aggregated iteratively to compute a globally optimal model.

In this paper, we introduce a novel federated privacy-preserving tool for the identification of differentially expressed genes called *Flimma*. Our new tool represents a federated implementation of the popular differential expression analysis workflow *limma voom* [42], one of the standard pipelines widely applied in the field for expression analyses. We have chosen *limma voom* among other popular count-based methods, because it is comparably fast without sacrificing accuracy [5]. Besides that, *Flimma* could be easily modified for handling microarray data, since the *limma* method was originally designed for such data [43], and only later extended to RNA-seq data via *voom [5]*.

Our *Flimma* implementation provides several advantages over the existing approaches for gene expression analysis (Fig. 1). Unlike *limma voom*, *Flimma* by design preserves the privacy of the data in the cohorts since the expression profiles never leave the local execution sites. In contrast to meta-analysis approaches, *Flimma* is particularly robust against heterogeneous distributions of data (in particular of confounders and class labels) across the different cohorts, which makes it a powerful alternative for multi-center studies where patient privacy is a key concern.

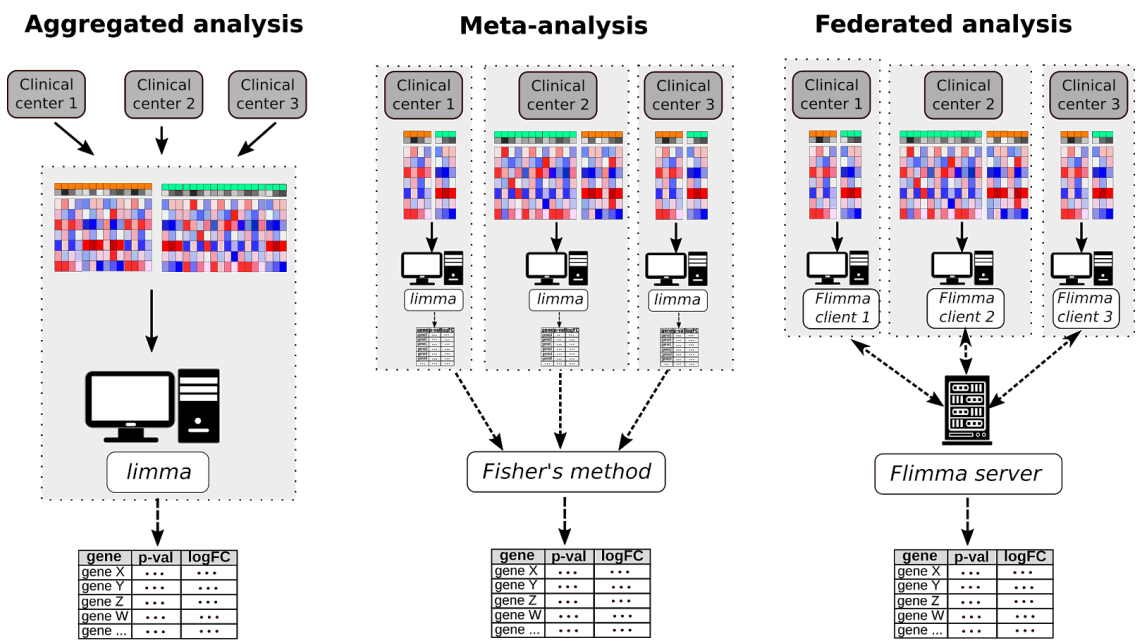

**Figure 1.** Gene expression analysis in case of multi-center studies. Bold arrows show the exchange of raw data, dashed arrows — the exchange of model parameters or summary statistics.

## Methods

### The *limma voom* workflow

*limma voom* is the state-of-the-art method for differential expression analysis. Initially designed for microarrays [43], it was extended by the *voom* function, which removes the mean-variance trend from RNA-seq data and makes it suitable for analysis by *limma [5]*. Recently, the authors of *limma* published an updated guideline on the recommended *limma voom* workflow [42]. Data preprocessing steps of this workflow include removal of weakly expressed genes using the *filterByExpr* function from the edgeR package, conversion of raw read counts to log2-transformed counts per million (log-CPM), and normalization of gene expression distributions. We only differ from this workflow by using the upper-quartile (UQ) normalization [35] instead of the trimmed mean of M-values (TMM) normalization [44], since the latter would require disclosing one of the sample profiles to all participants. Although UQ is not the only normalization method that could be implemented in a federated fashion, we have chosen it because it is one of the most widely used in the field [45,46]. Since no normalization method outperforms others in all cases [47,48], we are going to implement more federated normalization methods in the future. Furthermore, given the matrix of normalized log-CPM values and the design matrix, *voom* computes precision weights, which compensate for the mean-variance bias that is typical for RNA-seq data and thus makes them suitable for use in *limma*.

## *Flimma*

### Workflow

*Flimma* implements a federated version of the *limma voom* workflow, allowing privacy-preserving detection of differentially expressed genes. The scheme of the *Flimma* workflow is presented in Figure 2.

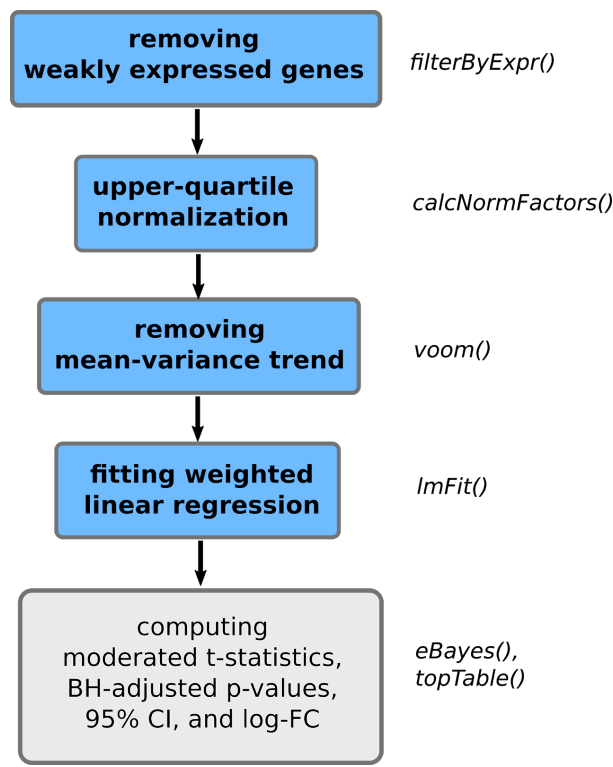

**Figure 2.** The scheme of *Flimma* workflow. Steps that were reimplemented in a federated fashion are shown in blue. The names of the functions used in the *limma voom* workflow are shown on the right of the flowchart.

First, genes that do not have sufficient counts for further statistical analysis are removed. For this, we implemented a federated version of the *filterByExprs* function [49] from the *edgeR* package, which employs two filters: *min_total_count* filter and CPM cutoff. The first filter removes genes whose sum of counts over all samples does not exceed *min_total_count* threshold. The second filter excludes genes expressed in insufficient number of samples. It keeps only genes where at least *min_n_samples* samples pass the CPM cutoff. This cutoff is calculated as a ratio of *min_count* over the median library size multiplied by $10^6$, where *min_n_samples* is defined by the smallest group size in the design matrix. The function parameters *min_count* and *min_total_count* are set to 10 and 15 by default and can be adjusted by the user.

UQ normalization performed in the second step of the pipeline requires the exchange of scaled normalization factors which cannot be used to reveal any private data. The third and the fourth steps of the workflow resemble the *voom* and *lmFit* functions from the *limma* package, which are fitting linear regression models. For training the linear regression model in the federated fashion, *Flimma* utilizes the same approach described by [50]. For each gene, each of $n$ clients computes $(X^i)^T X^i$ and $(X^i)^T Y^i$, where $X^i$ is the design matrix, $Y^i$ is the vector of normalized log2-CPM values for the gene, and $i$ is the index of a client, and sends them to the server. The server computes global $X^T X$, and $X^T Y$, $\beta$, and unscaled standard errors of the coefficients:

$$X^T X = \sum_{i=1}^{C} (X^i)^T X^i$$
$$X^T Y = \sum_{i=1}^{C} (X^i)^T Y^i$$
$$\beta = (X^T X)^{-1} X^T Y$$
$$uSE_\beta = diag(X^T X)$$

Global coefficients $\beta$ are sent back to the clients, which locally compute fitted log-CPM
$$\hat{Y}^i = X^i \beta,$$
and the sums of squared errors
$$SSE^i = \sum_{s=1}^{m^i} (y_s^i - \hat{y}_s^i)^2,$$
where $s$ is sample index and $m^i$ is the total number of samples in the $i$-th client.

The server collects $SSE^i$ from clients and computes estimated residual standard deviations for each gene:
$$\sigma = \sqrt{\frac{\sum_i SSE^i}{(\sum_i m^i) - k}}$$

The fifth step involves only $\beta$, $\sigma^2$, and unscaled standard errors, and therefore, does not require to be federated. All subsequent computations are performed on the server-side in the same way as done by the original *limma voom*.

Implementation

*Flimma* adopts the implementation of *sPLINK* [36] and consists of three software components: a client, a server, and a web interface. The coordinating user (coordinator) creates a project, defines the study parameters (e.g. confounding factors, etc.), and invites the distributed participants using the web interface. Each participant executes the client package locally to join the study with a token received from the coordinator and to select the local dataset to contribute to the study. The server component can run in a separate machine, at one of the clients, or at the coordinator's machine to orchestrate the clients and to aggregate their local model parameters, and to provide the global model to the coordinator.

As the original *limma voom*, each *Flimma* client accepts a matrix of read counts and a design matrix, specifying class labels and covariates for each sample. *Flimma* outputs a table with p-values, fold-changes and moderated t statistics for each gene.

*Flimma* is publicly available at https://exbio.wzw.tum.de/flimma/. The "HowTo" page provides a quick-start guide for *Flimma* along with test data and describes input file formats.

## Meta-analysis approaches

Three classes of meta-analysis approaches can be distinguished: effect size combination methods, p-value combination methods, and non-parametric methods [33]. Effect size combination methods estimate variances of effect sizes for every gene and compute global effect sizes as a weighted sum of local effect sizes divided by the sum of all weights. This class includes the fixed effects model (FEM) and the random effects model (REM), which differ in the way they compute weights [30]. FEM calculates the weights as the inverse of the within-study variance. REM assumes that total variance includes within-study and between-study variance components and calculates the inverse of their sum. Both methods calculate p-values given global effect sizes and assuming their normal distribution. We chose REM since it is more robust to data heterogeneity than FEM and more widely used [51].

P-value combination methods are based on the assumption that the sum, minimum or maximum of log-transformed p-values obtained in independent studies follow a certain distribution [33]. These methods are thought to be more suitable for imbalanced scenarios than effect size combination methods [52]. From this class of methods, we chose Fisher's method [27] because it is most sensitive to small p-values [53], and Stouffer's method (also known as z-method) [28] since it was shown to be superior to Fisher's method in some cases [54].

Non-parametric rank-based methods estimate global permutation-based p-value, by comparing the sum or the product of ranks obtained for the observed matrix of ranks with the same summary statistics calculated on shuffled rank matrices. Although the Rank Product method [29] is much more computationally expensive than the Rank Sum, the first gives more robust results [55].

In this work, we used the REM and Fisher's method from *metaVolcanoR* package [56], the implementation of Stouffer's method from *MetaDE* package [57] and *RankProd* package [58] for Rank Product method. For all selected meta-analysis methods except REM, global fold-change was calculated as a mean of local fold-changes.

## Evaluation

The main result of differential expression analysis is a list of genes with p-values and log-fold-changes, reflecting the significance and the strength of differential expression, respectively. To validate the results of *Flimma* and demonstrate its advantage over meta-analysis approaches, we compared the *Flimma* and meta-analysis results obtained on artificial dataset splits to the results of *limma voom* applied on the aggregated datasets.

We chose two large datasets comprising RNA-seq gene expression profiles of human-derived samples. The first dataset included 850 expression profiles of human breast tumors from TCGA-BRCA cohort [59], classified as luminal or basal subtypes and annotated with patient age and tumor stage. We searched for genes differentially expressed between luminal and basal subtypes and included the age of diagnosis and tumor stage as covariates. The second dataset comprised 1277 skin expression profiles from GTEx [60] with sun exposure as target class label and patient age and sex as covariates. Each dataset has been divided into cohorts to model the multi-party setting under various scenarios (see Datasets section for details).

In all tests, we applied *limma voom* on the complete dataset and on each of its partitions independently. The p-values and effect sizes computed by *limma voom* on the aggregated datasets were treated as ground truth, and those obtained on cohorts were used as input for the meta-analysis methods, which aggregated them to the global p-values.

To avoid manual execution of *Flimma* GUI for every test, we used a script performing exactly the same computations as the web version of *Flimma*. The code for running *Flimma* and its baselines, and the instructions for data download and preprocessing are available at https://github.com/ozolotareva/Flimma.

For each method, we considered a gene determined as differentially expressed, if it has |log(FC)|>1 and BH-adjusted p-value < 0.05. For the results produced by each method, we computed the RMSE, the precision, the recall, the F1 score, the Pearson and the Spearman correlation. Since only a small number of the most significantly differentially expressed genes is of interest for some research tasks, we have also investigated how the performance of the methods varies with the numbers of top-ranked genes selected.

## Datasets

### TCGA breast cancer data

Unprocessed read counts summarized to gene-level and clinical annotations of samples were downloaded from http://gdac.broadinstitute.org/runs/stddata__2016_01_28/data/BRCA/20160128. 850 expression profiles classified as luminal, or basal-like subtypes and annotated with the age of diagnosis and tumor stage were kept. Although breast cancer samples are classified into

4-6 subtypes [61–63], we focused on the most frequent subtypes for evaluation purposes. Luminal and basal subtypes are well distinguishable at the level of gene expression [59,61] (Supplementary Fig. S1A). We searched for genes differentially expressed between these subtypes and included the age of diagnosis and tumor stage as covariates. The luminal subtype is subdivided into luminal A (LumA) and luminal B (LumB) subtypes [62]. However, the LumA subtype was not included in the model as a covariate and we modeled the presence of an unknown disease subtype in our experiments.

### GTEx skin data

Raw read counts per gene were obtained from the GTEx v8 portal website (https://www.gtexportal.org/home/datasets). Expression profiles of sun-exposed and non-sun-exposed skin samples annotated with mean ischemic time and sex were kept. The resulting dataset comprises 1277 expression profiles of 677 sun-exposed and 600 non-sun-exposed skin samples, also annotated with sex and ischemic time. In contrast to the TCGA-BRCA dataset, a smaller fraction of genes was differentially expressed between sun-exposed and non-exposed skin samples (Supplementary Fig. S1A). Besides patient age and sex, samples were annotated with ischemic time, i.e. the time between patient death or sample withdrawal and sample fixation, or freezing. Ischemic time was not included in linear models but varied between cohorts in imbalanced scenarios, thus serving as an unknown confounder related to differences in sample preprocessing.

### Generation of artificially distributed and heterogeneous datasets

To demonstrate the robustness of *Flimma,* we split both datasets differently in a balanced, a mildly imbalanced, and a strongly imbalanced scenario. In the balanced scenario, each sample was randomly assigned to one of three equal-sized cohorts with a similar distribution of covariates. In the imbalanced scenarios, the fractions of target classes and the distributions of some covariates differed among cohorts. Cohort sizes were unequal and related as 1:2:4 and 1:3:9 for the mildly and the strongly imbalanced scenarios, respectively. In the TCGA-BRCA dataset, we introduced an imbalance of luminal and basal subtype frequencies and, in addition, changed the frequency of the LumA subtype (Table 1). In the GTEx skin dataset, the fraction of sun-exposed skin samples and the median of mean ischemic times were made unequal between cohorts in imbalanced scenarios (Table 2).

|  | cohort sizes | | | frequency of Basal subtype | | | frequency of LumA subtype | | |
|---|---|---|---|---|---|---|---|---|---|
|  | *Cohort 1* | *Cohort 2* | *Cohort 3* | *Cohort 1* | *Cohort 2* | *Cohort 3* | *Cohort 1* | *Cohort 2* | *Cohort 3* |
| **Balanced** | 283 | 283 | 284 | 0.20 | 0.20 | 0.20 | 0.57 | 0.57 | 0.58 |
| **Mildly imbalanced** | 121 | 242 | 487 | 0.10 | 0.30 | 0.17 | 0.40 | 0.50 | 0.66 |
| **Strongly imbalanced** | 65 | 196 | 589 | 0.25 | 0.50 | 0.09 | 0.14 | 0.50 | 0.65 |

**Table 1.** Characteristics of three scenarios for the TCGA-BRCA dataset. The distributions of ages and tumor stages were balanced.

|  | cohort sizes | | | fraction of sun-exposed skin samples | | | Mean ischemic time (minutes) | | |
|---|---|---|---|---|---|---|---|---|---|
|  | *Cohort 1* | *Cohort 2* | *Cohort 3* | *Cohort 1* | *Cohort 2* | *Cohort 3* | *Cohort 1* | *Cohort 2* | *Cohort 3* |
| **Balanced** | 425 | 425 | 427 | 0.53 | 0.53 | 0.53 | 629 | 636 | 636 |
| **Mildly imbalanced** | 181 | 363 | 733 | 0.4 | 0.65 | 0.51 | 490 | 620 | 676 |
| **Strongly imbalanced** | 97 | 293 | 887 | 0.8 | 0.4 | 0.54 | 347 | 646 | 661 |

**Table 2.** Characteristics of the scenarios for the GTEx skin dataset. The frequencies of samples obtained from male and female individuals were similar in all cohorts (between 30% and 34% samples from females in all scenarios).

## Results

We applied *Flimma* and four meta-analyses approaches on two real-world datasets: a breast cancer expression dataset from TCGA [59], and a skin dataset from GTEx [60]. To assess *Flimma*'s power, we model the multi-party setting by randomly partitioning both datasets into virtual cohorts, while introducing different levels of imbalance w.r.t. target class labels and covariate distributions. We employed two realistic scenarios leading to heterogeneity of sample distribution between local cohorts. First, we split the breast cancer dataset such that the (virtual) cohorts yield different frequencies of the LumA subtype to simulate an imbalanced distribution of disease subtypes collected at different clinical centers. In addition, we partitioned it according to tissue source sites. Second, we split the GTEx skin dataset by the mean ischemic time to illustrate the effect of potential confounders such as differences in sample collection and/or processing between the participating laboratories.

We then compared *Flimma* with popular meta-analysis tools using the *limma voom* results on the full data sets as gold standard. In summary, *Flimma* obtained the same results as *limma voom* in all tests. Across all experiments, the maximal absolute difference for log-transformed p-values and log-fold-change values computed by *Flimma* and *limma voom* did not exceed 9.30e-2 (Supplementary Table S1). In contrast, the results of the meta-analysis methods diverged from the results of *limma voom*, and this effect was especially pronounced in imbalanced scenarios.

## Evaluation on artificial dataset splits

We compared negative log-transformed p-values computed by all privacy-preserving approaches (i.e., *Flimma* and meta-analysis methods) with the results obtained by running *limma voom* on the combined dataset. For the privacy-preserving approaches, we computed the root mean square error (RMSE), the precision, the recall, the F1 score, the Pearson and the Spearman correlation w.r.t. the results of the aggregated analysis with *limma voom*, which we treated as ground truth.

As shown in Fig. 3, Tables 3-4, and Supplementary Tables S2-S3, *Flimma* produces the same p-values as the aggregated analysis with *limma voom* in all scenarios, including the imbalanced ones. This implies that *Flimma* is robust against heterogeneous data distributions across the clients. However, this is not the case for the meta-analysis approaches. In general, their RMSEs increase (and Pearson correlations decrease) as the scenarios become more imbalanced, and they introduce false positives and false negatives even in the balanced scenario. In spite of the difference in p-values calculated by all meta-analysis methods, their gene rankings were quite similar to the ranking produced by the aggregated *limma voom* (the Spearman correlation varied between 0.74 to 0.99 in all experiments).

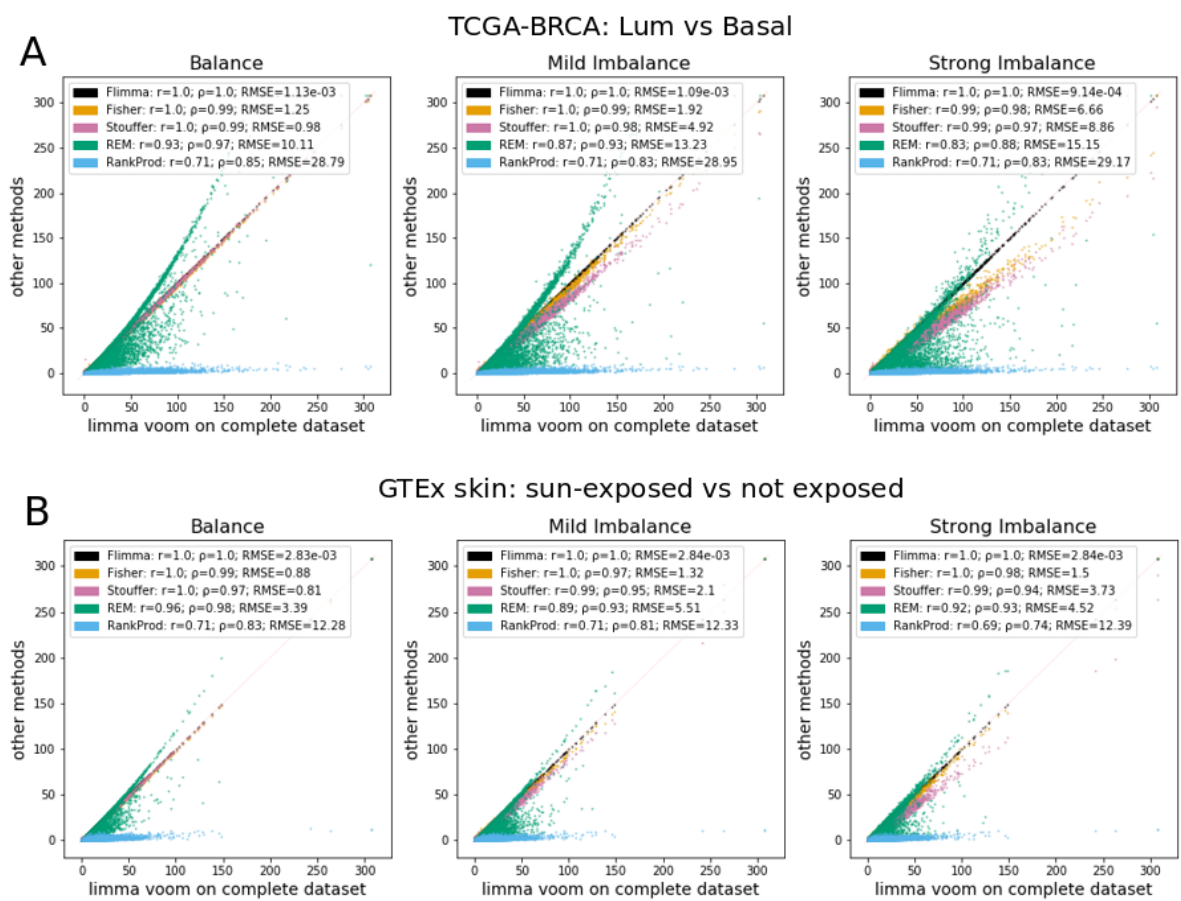

**Figure 3**. The comparison of negative log-transformed p-values computed by *Flimma* and meta-analysis methods (y-axis) with p-values obtained by *limma* on the aggregated dataset (x-axis) in three scenarios on A. TCGA-BRCA and B. GTEx skin datasets. Pearson correlation coefficient (r), Spearman correlation coefficient (ρ), and root-mean squared error (RMSE) calculated for each method are reported in the legend.

| metric | F1 | | | FP | | | FN | | |
|---|---|---|---|---|---|---|---|---|---|
| Scenario | Balanced | Mildly Imbalanced | Strongly Imbalanced | Balanced | Mildly Imbalanced | Strongly Imbalanced | Balanced | Mildly Imbalanced | Strongly Imbalanced |
| ***Flimma*** | <u>1.00</u> | <u>1.00</u> | <u>1.00</u> | <u>0</u> | <u>0</u> | <u>0</u> | <u>0</u> | <u>0</u> | <u>0</u> |
| ***Fisher*** | <u>1.00</u> | 0.92 | 0.93 | 14 | 248 | 192 | 8 | 290 | 265 |
| ***Stouffer*** | <u>1.00</u> | 0.92 | 0.93 | 14 | 245 | 189 | 9 | 290 | 265 |
| ***REM*** | <u>1.00</u> | 0.97 | 0.95 | 12 | 80 | 121 | 17 | 119 | 215 |
| ***RankProd*** | <u>1.00</u> | 0.92 | 0.93 | 14 | 243 | 193 | 12 | 295 | 274 |

**Table 3.** F1 score, the number of false positives (FP) and the number of false negatives (FN) obtained on TCGA-BRCA dataset in three scenarios. Values corresponding the best performance over all methods are underlined. All calculated performance measures are reported in Supplementary Table S2.

| metric | F1 | | | FP | | | FN | | |
|---|---|---|---|---|---|---|---|---|---|
| Scenario | Balanced | Mildly Imbalanced | Strongly Imbalanced | Balanced | Mildly Imbalanced | Strongly Imbalanced | Balanced | Mildly Imbalanced | Strongly Imbalanced |
| ***Flimma*** | <u>1.00</u> | <u>1.00</u> | <u>1.00</u> | <u>0</u> | <u>0</u> | <u>0</u> | <u>0</u> | <u>0</u> | <u>0</u> |
| ***Fisher*** | 0.99 | 0.91 | 0.83 | 4 | 32 | 67 | <u>0</u> | 18 | 33 |
| ***Stouffer*** | 0.99 | 0.91 | 0.83 | 4 | 32 | 67 | <u>0</u> | 18 | 33 |
| ***REM*** | 0.99 | 0.95 | 0.94 | 4 | 15 | 21 | 2 | 14 | 12 |
| ***RankProd*** | 0.99 | 0.91 | 0.83 | 4 | 32 | 67 | <u>0</u> | 18 | 33 |

**Table 4.** F1 score, the number of false positives (FP) and the number of false negatives (FN) obtained on GTEx skin dataset in three scenarios. Values corresponding the best performance over all methods are underlined. All calculated performance measures are reported in Supplementary Table S3.

## Performance for top-ranked genes

Since some research tasks such as biomarker discovery require the identification of a small number of significantly differentially expressed genes, we investigated how the performance of the methods varies with altered numbers of selected top differentially expressed genes after sorting by p-value (Fig. 4 and Supplementary Figures S2-3). Again, *Flimma* perfectly reproduced the results of aggregated *limma voom* in all scenarios and outperformed all meta-analysis approaches. Fisher's and Stouffer's methods demonstrated almost perfect performance in the balanced scenario, but their performance decreased in the imbalanced ones.

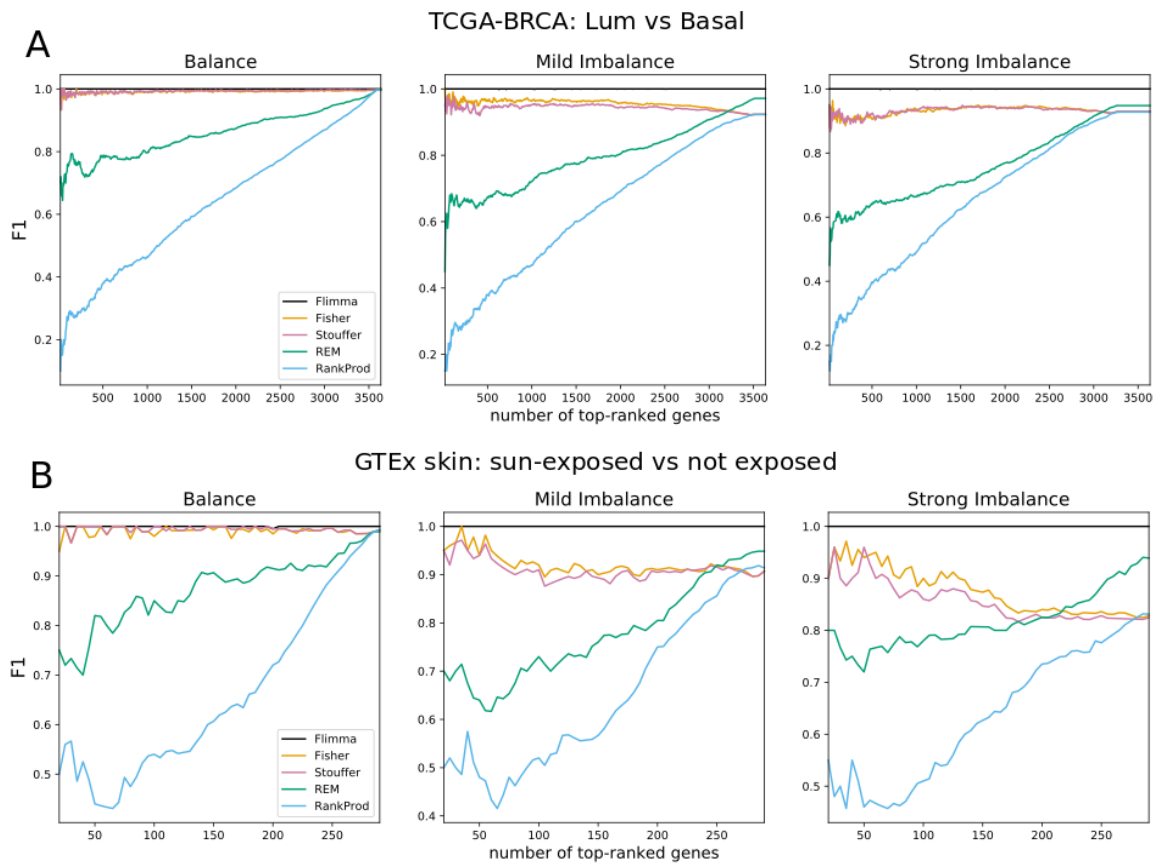

**Figure 4.** The dependency of the F1 score on the number of top-ranked genes considered to be differentially expressed. Genes were ranked in order of their negative log-transformed p-values decreasing and the number of top-ranked genes varied between 20 and 3500 (for TCGA-BRCA dataset, A) and 300 (for GTEx Skin dataset, B) with step 5.

## Splitting TCGA-BRCA by sample source site

TCGA is a multi-center project and tumor samples of TCGA-BRCA datasets were collected at 37 different clinical centers, which can result in some between-center variability. Therefore, we also evaluated *Flimma* and its baselines on a more realistic scenario, where TCGA-BRCA dataset was split according to the sample source sites, but we kept only 14 of the 37 cohorts, such that each cohort contained at least 3 samples of LumA and basal subtype.

We selected 3, 5, 7, 10, and 14 cohorts such that subtype frequencies, mean stage, and age are dissimilar across the selected cohorts (cf. Supplementary Table S4 for details). We also added additional terms in linear models to account for possible cohort effects. Similar to the previous experiments, *Flimma* clearly outperforms all meta-analysis approaches in terms of RMSE, precision, and recall (Table 5 and Supplementary Table S5).

| metric | RMSE | | | | | Precision | | | | | Recall | | | | |
|---|---|---|---|---|---|---|---|---|---|---|---|---|---|---|---|
| The number of cohorts | 3 | 5 | 7 | 10 | 14 | 3 | 5 | 7 | 10 | 14 | 3 | 5 | 7 | 10 | 14 |
| ***Flimma*** | <u>0.0008</u> | <u>0.0007</u> | <u>0.0008</u> | <u>0.0017</u> | <u>0.0012</u> | <u>1.00</u> | <u>1.00</u> | <u>1.00</u> | <u>1.00</u> | <u>1.00</u> | <u>1.00</u> | <u>1.00</u> | <u>1.00</u> | <u>1.00</u> | <u>1.00</u> |
| ***Fisher*** | 0.94 | 1.82 | 2.53 | 3.86 | 5.37 | 0.85 | 0.88 | 0.90 | 0.93 | 0.95 | 0.92 | 0.95 | 0.95 | 0.96 | 0.97 |
| ***Stouffer*** | 1.47 | 2.21 | 2.87 | 4.26 | 5.68 | 0.85 | 0.88 | 0.91 | 0.93 | 0.95 | 0.89 | 0.93 | 0.94 | 0.96 | 0.97 |
| ***REM*** | 2.73 | 3.68 | 4.75 | 7.21 | 8.50 | 0.93 | 0.94 | 0.95 | 0.97 | 0.97 | 0.93 | 0.96 | 0.97 | 0.98 | 0.98 |
| ***RankProd*** | 5.16 | 8.19 | 11.32 | 18.92 | 23.50 | 0.92 | 0.87 | 0.90 | 0.93 | 0.95 | 0.87 | 0.96 | 0.96 | 0.96 | 0.97 |

**Table 5.** RMSE, precision, and recall obtained by Flimma and the meta-analysis tools on TCGA-BRCA datasets split by tissue source sites. Values corresponding the best performance over all methods are underlined.

## Discussion

In this work, we presented *Flimma*, the first privacy-preserving method for differential expression analysis. While *Flimma* results are mathematically equivalent to *limma voom*, *Flimma* can operate on distributed cohorts without the disclosure of sensitive data. In this work, we have demonstrated that *Flimma* is superior to meta-analysis in imbalanced scenarios when the distributions of class labels or covariates are not identical between cohorts.

One limitation of this work is the absence of a gold standard for the evaluation of differential expression analysis results. ABCD mixtures used in RNA-seq benchmark projects [1,64] are not suitable for this study, since only five or less replicates of each mixture are sequenced by each participant. Although these projects are multi-center studies, such a small number of samples per participating center would not be realistic for mimicking modern biomedical studies involving human patients. Moreover, with these artificial mixtures, we could not model biological variability which is intrinsic of real-world patient-derived data. Therefore, we have chosen two large and well-annotated patient-derived expression datasets, split them into parts modeling independent cohorts, and considered the results of *limma voom* obtained on the combined datasets as ground truth.

Another limitation of this work is that we did not investigate the effect of sequencing platforms and pipelines used to quantify read counts. We assumed that all participants utilized the same experimental protocols and *in silico* pipelines to generate read count matrices, but in reality this may not be the case. To overcome this limitation, the development of a federated version of batch effect removal methods, such as ComBat [65,66] or RUV [45], is necessary.

While *limma voom* is a state-of-the-art method for differential expression analysis that performs favoroubly in benchmarks [5], other methods for normalization (e.g. quantile normalization [67]) and differential expression analysis (such as edgeR [3], DESeq2 [6], or sleuth [9]) exist and may yield different results depending on the dataset used. We thus consider to extend *Flimma* with federated implementations of alternative methods in the future.

## Conclusions

*Flimma* is a privacy-by-design tool for the federated identification of differentially expressed genes. Unlike *limma voom, Flimma* preserves the privacy of the data in the cohorts as raw data never has to leave a local hospital. In contrast to meta-analysis approaches, *Flimma* is robust against heterogeneous distribution of data across the different sites. In summary, *Flimma* is a perfect alternative for multi-center gene expression projects, as it preserves patient privacy by design while providing the same power as a centralized analysis. It is user friendly and publicly available at https://exbio.wzw.tum.de/flimma/ including tutorials and a video documentation on its principle and application to real data.

## Competing interests

The authors declare that they have no competing interests.

## Author's contributions

OZ, ML, and JB conceived and designed the study. OZ preprocessed the data, performed the experiments and analyzed the results. OZ and RN developed the federated algorithms. RN implemented the client and server components. JM implemented the web interface. All authors provided critical feedback and helped in the interpretation of data, manuscript writing, and approved the final version.

## Funding

This work was supported by the German Federal Ministry of Education and Research (BMBF) within the framework of "CLINSPECT-M" (grant 031L0214A) and within the framework of the *e:Med *research and funding concept "Sys_CARE" (*grant 01ZX1908A*). This project has received funding from the European Union's Horizon2020 research and innovation programme under grant agreement No 826078 (FeatureCloud) and No 777111 (REPO-TRIAL). This publication reflects only the authors' view and the European Commission is not responsible for any use that may be made of the information it contains. JB was partially funded by his VILLUM Young Investigator Grant No 13154.

# Supplementary Figures and Tables

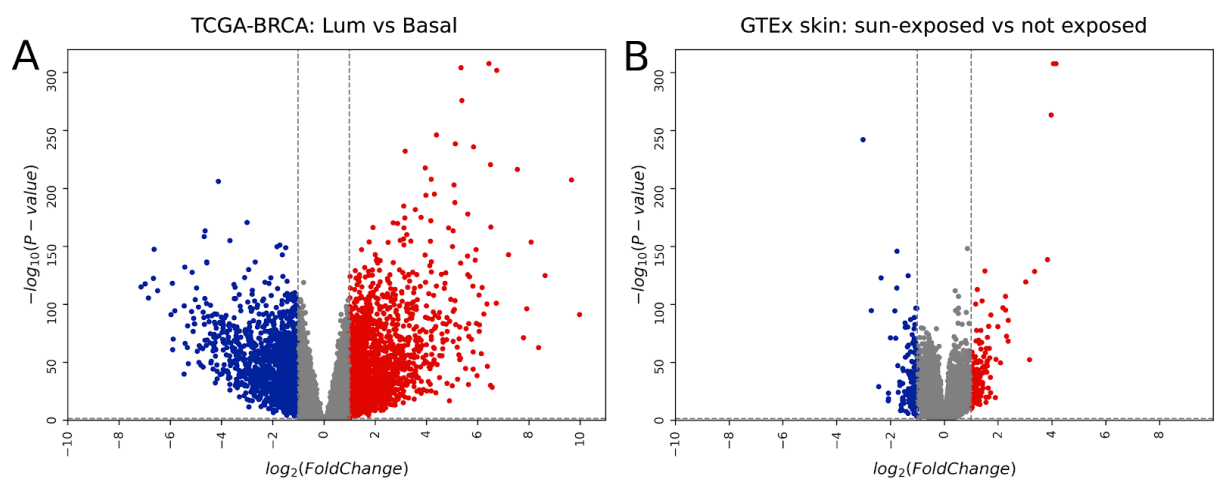

**Figure S1.** The results of differential expression analysis performed by *limma* on the aggregated TCGA-BRCA (A) and GTEx Skin datasets (B). 3635 and 288 genes were differentially expressed with absolute log-fold change |logFC| > 1 and BH-adjusted p-value < 0.05 according to *limma voom* applied on pooled TCGA-BRCA and GTEx Skin datasets respectively.

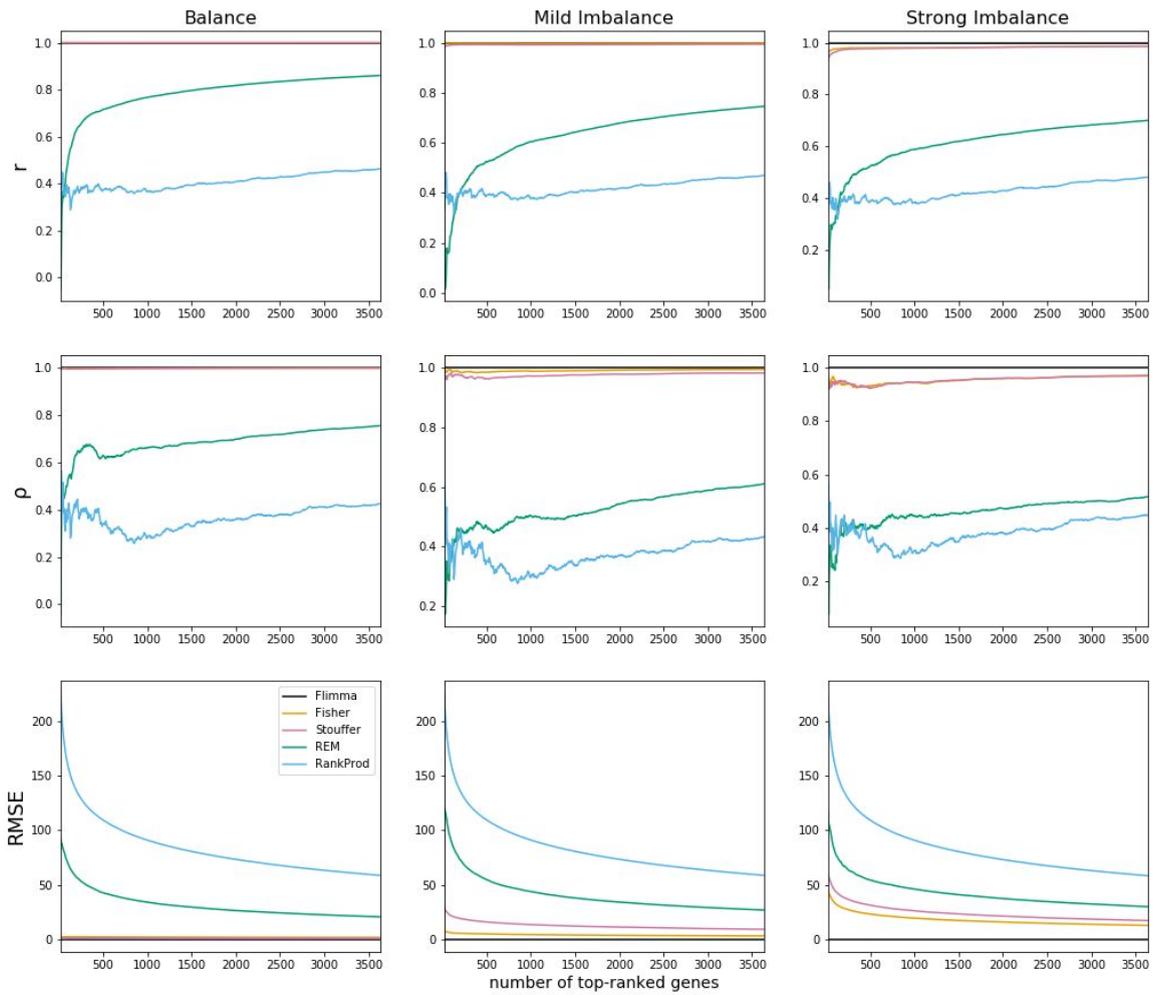

**Figure S2.** The dependency of Pearson's and Spearman's correlation coefficients and RMSE on the number of top-ranked genes considered to be differentially expressed obtained on TCGA-BRCA dataset.

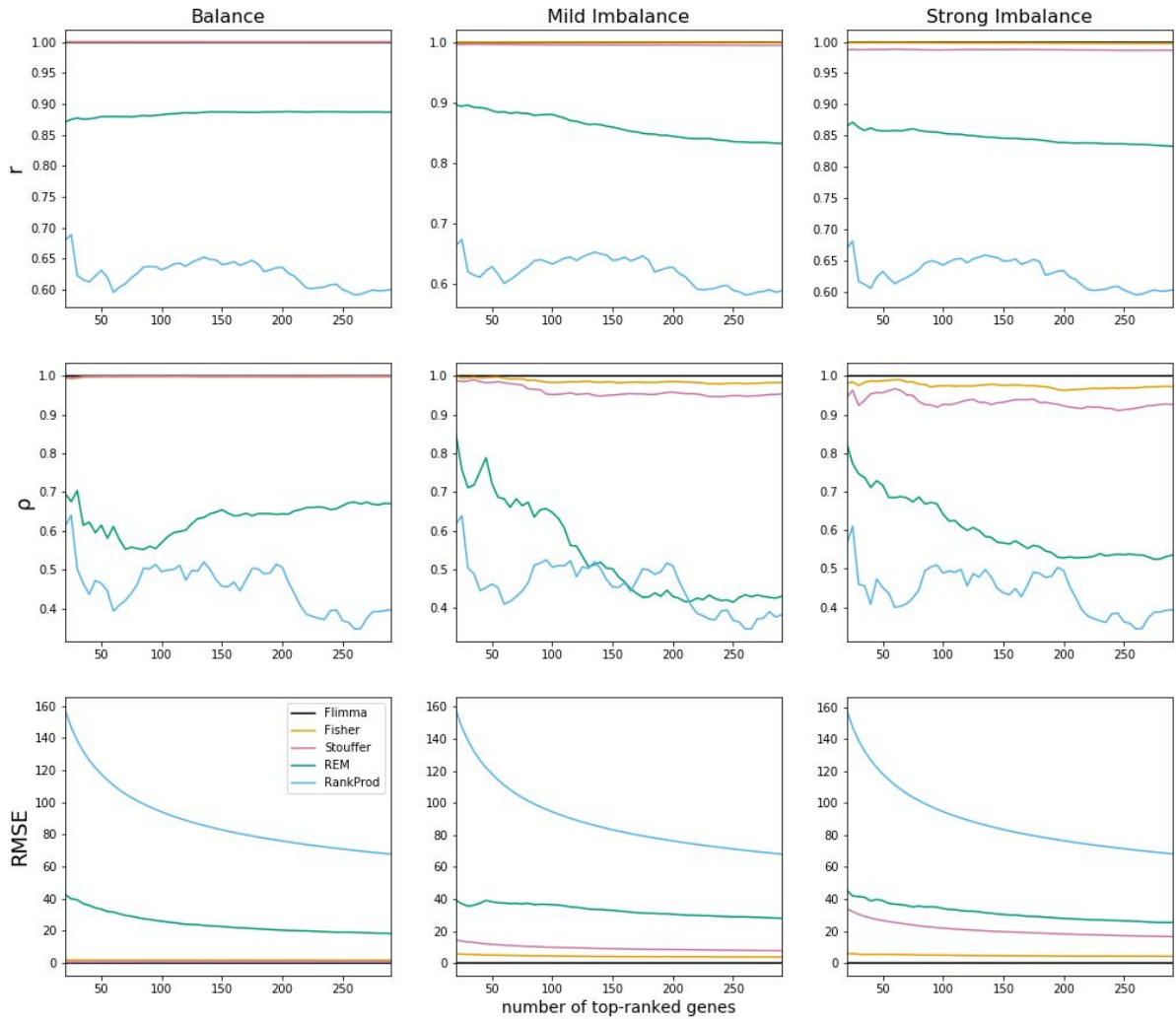

**Figure S3.** The dependency of Pearson's and Spearman's correlation coefficients and RMSE on the number of top-ranked genes considered to be differentially expressed obtained on GTEx skin dataset.

Supplementary Table S1. The difference between the results of *limma voom* and *Flimma*.

| | | absolute difference of -log10(p-values) | | | absolute difference of log2(fold-chages) | | |
|---|---|---|---|---|---|---|---|
| | | min | mean | max | min | mean | max |
| TCGA-BRCA luminal vs basal subtype | Balanced | 0.00E+00 | 3.86E-04 | 7.21E-02 | 2.80E-11 | 1.53E-05 | 1.15E-03 |
| | Mildly Imbalanced | 0.00E+00 | 3.69E-04 | 7.21E-02 | 2.80E-11 | 1.37E-05 | 1.15E-03 |
| | Strongly Imbalanced | 0.00E+00 | 3.22E-04 | 7.21E-02 | 2.81E-11 | 1.09E-05 | 6.63E-04 |
| | 3 cohorts | 6.78E-09 | 2.02E-04 | 2.98E-02 | 7.45E-12 | 3.65E-05 | 7.23E-03 |
| | 5 cohorts | 1.88E-08 | 2.11E-04 | 2.65E-02 | 3.11E-10 | 2.40E-05 | 2.29E-03 |
| | 7 cohorts | 2.52E-09 | 2.41E-04 | 5.49E-02 | 7.41E-10 | 2.00E-05 | 1.89E-03 |
| | 10 cohorts | 1.82E-08 | 4.19E-04 | 6.83E-02 | 4.30E-11 | 2.32E-05 | 2.95E-03 |
| | 14 cohorts | 0.00E+00 | 3.76E-04 | 9.30E-02 | 2.26E-10 | 1.45E-05 | 1.01E-03 |
| GTEx skin sun-exposed vs not sun-exposed | Balanced | 0.00E+00 | 1.73E-03 | 8.50E-02 | 1.54E-09 | 2.93E-05 | 1.13E-03 |
| | Mildly Imbalanced | 0.00E+00 | 1.73E-03 | 8.50E-02 | 1.54E-09 | 2.94E-05 | 1.13E-03 |
| | Strongly Imbalanced | 0.00E+00 | 1.73E-03 | 8.50E-02 | 1.54E-09 | 3.00E-05 | 1.33E-03 |

Supplementary Table S2. Performance on TCGA-BRCA dataset

| | Scenario | *Flimma* | *Fisher* | *Stouffer* | *REM* | *RankProd* |
|---|---|---|---|---|---|---|
| **FP** | Balanced | 0 | 14 | 14 | 12 | 14 |
| | Mildly Imbalanced | 0 | 248 | 245 | 80 | 243 |
| | Strongly Imbalanced | 0 | 192 | 189 | 121 | 193 |
| **FN** | Balanced | 0 | 8 | 9 | 17 | 12 |
| | Mildly Imbalanced | 0 | 290 | 290 | 119 | 295 |
| | Strongly Imbalanced | 0 | 265 | 265 | 215 | 274 |
| **F1** | Balanced | 1.00 | 1.00 | 1.00 | 1.00 | 1.00 |
| | Mildly Imbalanced | 1.00 | 0.92 | 0.92 | 0.97 | 0.92 |
| | Strongly Imbalanced | 1.00 | 0.93 | 0.93 | 0.95 | 0.93 |
| **Precision** | Balanced | 1.00 | 1.00 | 1.00 | 1.00 | 1.00 |
| | Mildly Imbalanced | 1.00 | 0.93 | 0.93 | 0.98 | 0.93 |
| | Strongly Imbalanced | 1.00 | 0.94 | 0.94 | 0.96 | 0.94 |
| **Recall** | Balanced | 1.00 | 1.00 | 1.00 | 1.00 | 1.00 |
| | Mildly Imbalanced | 1.00 | 0.92 | 0.92 | 0.97 | 0.92 |
| | Strongly Imbalanced | 1.00 | 0.92 | 0.92 | 0.93 | 0.92 |
| **RMSE** | Balanced | 0.0011 | 1.25 | 0.98 | 10.11 | 28.79 |
| | Mildly Imbalanced | 0.0011 | 1.92 | 4.92 | 13.23 | 28.95 |
| | Strongly Imbalanced | 0.0009 | 6.66 | 8.86 | 15.15 | 29.17 |
| **r** | Balanced | 1.00 | 1.00 | 1.00 | 0.93 | 0.71 |
| | Mildly Imbalanced | 1.00 | 1.00 | 1.00 | 0.87 | 0.71 |
| | Strongly Imbalanced | 1.00 | 0.99 | 0.99 | 0.83 | 0.71 |
| **ρ** | Balanced | 1.00 | 0.99 | 0.99 | 0.97 | 0.85 |
| | Mildly Imbalanced | 1.00 | 0.99 | 0.98 | 0.93 | 0.83 |
| | Strongly Imbalanced | 1.00 | 0.98 | 0.97 | 0.88 | 0.83 |

Note: top header row groups columns as FP (Flimma, Fisher) and FN (Stouffer, REM, RankProd).

Supplementary Table S3. Performance on GTEx skin dataset.

|  | Scenario | *Flimma* | *Fisher* | *Stouffer* | *REM* | *RankProd* |
|---|---|---|---|---|---|---|
| **FP** | Balanced | 0 | 4 | 4 | 4 | 4 |
|  | Mildly Imbalanced | 0 | 32 | 32 | 15 | 32 |
|  | Strongly Imbalanced | 0 | 67 | 67 | 21 | 67 |
| **FN** | Balanced | 0 | 0 | 0 | 2 | 0 |
|  | Mildly Imbalanced | 0 | 18 | 18 | 14 | 18 |
|  | Strongly Imbalanced | 0 | 33 | 33 | 12 | 33 |
| **F1** | Balanced | 1.00 | 0.99 | 0.99 | 0.99 | 0.99 |
|  | Mildly Imbalanced | 1.00 | 0.91 | 0.91 | 0.95 | 0.91 |
|  | Strongly Imbalanced | 1.00 | 0.83 | 0.83 | 0.94 | 0.83 |
| **Precision** | Balanced | 1.00 | 0.99 | 0.99 | 0.99 | 0.99 |
|  | Mildly Imbalanced | 1.00 | 0.89 | 0.89 | 0.95 | 0.89 |
|  | Strongly Imbalanced | 1.00 | 0.79 | 0.79 | 0.93 | 0.79 |
| **Recall** | Balanced | 1.00 | 1.00 | 1.00 | 0.99 | 1.00 |
|  | Mildly Imbalanced | 1.00 | 0.94 | 0.94 | 0.95 | 0.94 |
|  | Strongly Imbalanced | 1.00 | 0.88 | 0.88 | 0.96 | 0.88 |
| **RMSE** | Balanced | 0.0028 | 0.88 | 0.81 | 3.39 | 12.28 |
|  | Mildly Imbalanced | 0.0028 | 1.32 | 2.10 | 5.51 | 12.33 |
|  | Strongly Imbalanced | 0.0028 | 1.50 | 3.73 | 4.52 | 12.39 |
| **r** | Balanced | 1.00 | 1.00 | 1.00 | 0.96 | 0.71 |
|  | Mildly Imbalanced | 1.00 | 1.00 | 0.99 | 0.89 | 0.71 |
|  | Strongly Imbalanced | 1.00 | 1.00 | 0.99 | 0.92 | 0.69 |
| **ρ** | Balanced | 1.00 | 0.99 | 0.97 | 0.98 | 0.83 |
|  | Mildly Imbalanced | 1.00 | 0.97 | 0.95 | 0.93 | 0.81 |
|  | Strongly Imbalanced | 1.00 | 0.98 | 0.94 | 0.93 | 0.74 |

Supplementary Table S4. TCGA-BRCA cohorts with at least 3 samples in LumA and Basal subtypes. 733 samples in 14 cohorts

|     | Lum | Basal | LumA | LumB | total samples | LumA fraction | Basal fraction | mean age | mean stage | 3 cohorts 199 samples | 5 cohorts 295 samples | 7 cohorts 383 samples | 10 cohorts 592 samples |
|-----|-----|-------|------|------|---------------|---------------|----------------|----------|------------|----------------------|----------------------|----------------------|-----------------------|
| BH  | 105 | 18    | 82   | 23   | 123           | 0.67          | 0.15           | 58.9     | 1.9        | +                    | +                    | +                    | +                     |
| A8  | 56  | 3     | 30   | 26   | 59            | 0.51          | 0.05           | 66.0     | 2.3        | +                    | +                    | +                    | +                     |
| OL  | 12  | 5     | 12   | 0    | 17            | 0.71          | 0.29           | 54.8     | 1.8        | +                    | +                    | +                    | +                     |
| D8  | 55  | 10    | 33   | 22   | 65            | 0.51          | 0.15           | 62.9     | 2.1        |                      | +                    | +                    | +                     |
| A7  | 18  | 13    | 14   | 4    | 31            | 0.45          | 0.42           | 59.6     | 2.1        |                      | +                    | +                    | +                     |
| AR  | 42  | 14    | 28   | 14   | 56            | 0.50          | 0.25           | 54.5     | 2.1        |                      |                      | +                    | +                     |
| C8  | 26  | 6     | 12   | 14   | 32            | 0.38          | 0.19           | 54.9     | 2.3        |                      |                      | +                    | +                     |
| E2  | 57  | 17    | 46   | 11   | 74            | 0.62          | 0.23           | 57.4     | 1.9        |                      |                      |                      | +                     |
| A2  | 58  | 21    | 40   | 18   | 79            | 0.51          | 0.27           | 57.3     | 2.1        |                      |                      |                      | +                     |
| E9  | 47  | 9     | 28   | 19   | 56            | 0.50          | 0.16           | 55.9     | 2.1        |                      |                      |                      | +                     |
| B6  | 30  | 10    | 24   | 6    | 40            | 0.60          | 0.25           | 55.5     | 2.2        |                      |                      |                      |                       |
| AN  | 28  | 8     | 21   | 7    | 36            | 0.58          | 0.22           | 59.1     | 2.3        |                      |                      |                      |                       |
| AO  | 29  | 7     | 19   | 10   | 36            | 0.53          | 0.19           | 54.9     | 2.1        |                      |                      |                      |                       |
| EW  | 24  | 5     | 17   | 7    | 29            | 0.59          | 0.17           | 55.3     | 2.0        |                      |                      |                      |                       |

Supplementary Table S5. Performance on TCGA-BRCA datasets split by tissue source sites.

| metric | dataset | *Flimma* | *Fisher* | *Stouffer* | *REM* | *RankProd* |
|---|---|---|---|---|---|---|
| **FP** | 3 cohorts | 0 | 600 | 565 | 252 | 290 |
| | 5 cohorts | 1 | 452 | 425 | 232 | 505 |
| | 7 cohorts | 0 | 352 | 335 | 178 | 381 |
| | 10 cohorts | 0 | 255 | 250 | 119 | 265 |
| | 14 cohorts | 0 | 188 | 177 | 102 | 190 |
| **FN** | 3 cohorts | 0 | 298 | 389 | 260 | 473 |
| | 5 cohorts | 0 | 189 | 245 | 130 | 136 |
| | 7 cohorts | 0 | 174 | 212 | 109 | 137 |
| | 10 cohorts | 0 | 138 | 152 | 72 | 130 |
| | 14 cohorts | 0 | 108 | 118 | 85 | 102 |
| **Precision** | 3 cohorts | 1.00 | 0.85 | 0.85 | 0.93 | 0.92 |
| | 5 cohorts | 1.00 | 0.88 | 0.88 | 0.94 | 0.87 |
| | 7 cohorts | 1.00 | 0.90 | 0.91 | 0.95 | 0.90 |
| | 10 cohorts | 1.00 | 0.93 | 0.93 | 0.97 | 0.93 |
| | 14 cohorts | 1.00 | 0.95 | 0.95 | 0.97 | 0.95 |
| **Recall** | 3 cohorts | 1.00 | 0.92 | 0.89 | 0.93 | 0.87 |
| | 5 cohorts | 1.00 | 0.95 | 0.93 | 0.96 | 0.96 |
| | 7 cohorts | 1.00 | 0.95 | 0.94 | 0.97 | 0.96 |
| | 10 cohorts | 1.00 | 0.96 | 0.96 | 0.98 | 0.96 |
| | 14 cohorts | 1.00 | 0.97 | 0.97 | 0.98 | 0.97 |
| **F1** | 3 cohorts | 1.00 | 0.88 | 0.87 | 0.93 | 0.89 |
| | 5 cohorts | 1.00 | 0.91 | 0.91 | 0.95 | 0.91 |
| | 7 cohorts | 1.00 | 0.93 | 0.92 | 0.96 | 0.93 |
| | 10 cohorts | 1.00 | 0.95 | 0.94 | 0.97 | 0.95 |
| | 14 cohorts | 1.00 | 0.96 | 0.96 | 0.97 | 0.96 |
| **RMSE** | 3 cohorts | 0.0008 | 0.94 | 1.47 | 2.73 | 5.16 |
| | 5 cohorts | 0.0007 | 1.82 | 2.21 | 3.68 | 8.19 |
| | 7 cohorts | 0.0008 | 2.53 | 2.87 | 4.75 | 11.32 |
| | 10 cohorts | 0.0017 | 3.86 | 4.26 | 7.21 | 18.92 |
| | 14 cohorts | 0.0012 | 5.37 | 5.68 | 8.50 | 23.50 |
| | 3 cohorts | 1.00 | 0.99 | 0.98 | 0.85 | 0.69 |
| **r** | 5 cohorts  | 1.00 | 0.99 | 0.99 | 0.90 | 0.70 |
|   | 7 cohorts  | 1.00 | 0.99 | 0.99 | 0.91 | 0.68 |
|   | 10 cohorts | 1.00 | 0.99 | 0.99 | 0.92 | 0.69 |
|   | 14 cohorts | 1.00 | 0.99 | 0.99 | 0.93 | 0.68 |
| **ρ** | 3 cohorts  | 1.00 | 0.96 | 0.91 | 0.94 | 0.80 |
|   | 5 cohorts  | 1.00 | 0.96 | 0.93 | 0.96 | 0.82 |
|   | 7 cohorts  | 1.00 | 0.96 | 0.94 | 0.97 | 0.82 |
|   | 10 cohorts | 1.00 | 0.97 | 0.96 | 0.98 | 0.83 |
|   | 14 cohorts | 1.00 | 0.97 | 0.96 | 0.98 | 0.83 |